\documentclass[pra,aps,twocolumn,superscriptaddress]{revtex4}
\usepackage{graphicx}
\usepackage{epsfig}
\usepackage{times}
\usepackage{amsmath}
\usepackage{amsfonts}
\usepackage{amssymb}
\usepackage{latexsym}
\usepackage[usenames]{color}
\usepackage{graphicx}
\usepackage{enumerate}
 \usepackage{times}
 \usepackage{algorithm}
 \usepackage{algorithmic}
 \usepackage{cases}

\begin{document}
\widetext
\title{\textcolor{black}{One-way} transfer of quantum states via decoherence}
\author{Yuichiro Matsuzaki}
\email{matsuzaki.yuichiro@lab.ntt.co.jp}
\affiliation{NTT Basic Research Laboratories, NTT Corporation, 3-1 Morinosato-Wakamiya, Atsugi, Kanagawa 243-0198, Japan}
\affiliation{NTT Theoretical Quantum Physics Center, NTT Corporation,
3-1 Morinosato-Wakamiya, Atsugi, Kanagawa 243-0198, Japan}
\author{Victor \textcolor{black}{M.} Bastidas}
\affiliation{NTT Basic Research Laboratories, NTT Corporation, 3-1 Morinosato-Wakamiya, Atsugi, Kanagawa 243-0198, Japan}
\affiliation{NTT Theoretical Quantum Physics Center, NTT Corporation,
3-1 Morinosato-Wakamiya, Atsugi, Kanagawa 243-0198, Japan}
\author{Yuki Takeuchi }
\affiliation{NTT Communication Science Laboratories, NTT Corporation, 3-1 Morinosato-Wakamiya, Atsugi, Kanagawa 243-0198, Japan}
\author{William J. Munro}
\affiliation{NTT Basic Research Laboratories, NTT Corporation, 3-1 Morinosato-Wakamiya, Atsugi, Kanagawa 243-0198, Japan}
\affiliation{NTT Theoretical Quantum Physics Center, NTT Corporation, 3-1 Morinosato-Wakamiya, Atsugi, Kanagawa 243-0198, Japan}
\affiliation{National Institute of Informatics, 2-1-2 Hitotsubashi, Chiyoda-ku, Tokyo 101-8430, Japan}
\author{Shiro Saito}
\affiliation{NTT Basic Research Laboratories, NTT Corporation, 3-1 Morinosato-Wakamiya, Atsugi, Kanagawa 243-0198, Japan}
\begin{abstract}
 \textcolor{black}{In many quantum information processing applications,
 it is important to be able to transfer a quantum state from one location to another - even within a local device.}
 Typical approaches \textcolor{black}{to implement the quantum state transfer}
 rely on unitary evolutions or
 measurement feedforward operations. However, these existing schemes require  accurate pulse operations
 and/or precise timing controls. Here, we propose a \textcolor{black}{one-way}
 transfer of the quantum state with near unit efficiency using dissipation from a tailored
 environment. After preparing an initial state, the transfer can be
 implemented without external time dependent operations.
 Moreover, our scheme is irreversible due to the non-unitary
 \textcolor{black}{evolution}, and so the \textcolor{black}{transferred} state remains in the
 same site once the system reaches \textcolor{black}{the} steady state. This is in stark contrast to the unitary
 state transfer where the quantum states continue to oscillate
 between different sites.
 Our novel quantum
 state transfer via the dissipation paves the way towards robust and
 practical quantum control. 
\end{abstract}
\maketitle

\section{Introduction}
Quantum state transfer is an essential technique to realize quantum
computation and quantum communication
\cite{cirac1997quantum,christandl2004perfect,sillanpaa2007coherent}.
One of typical approaches is to
use \textcolor{black}{flying qubits} such as optical photons
\cite{Cabrillo:1999p339,Bose:1999p326,liang2005realization,Lim:2005p364,Barrett:2005p363,Moehring:2007p337,MonroePRL,tey2008strong}.  
\textcolor{black}{In this case, a quantum state of a
stationary qubit can be transferred to the
flying photons using a solid state system such that they can interact with another
stationary qubit at a distant node.}
However, in such an
approach, 
the quantum
state transfer can be \textcolor{black}{typically} performed in a probabilistic way. 
Another approach is to use \textcolor{black}{a} stationary qubit array for
sending quantum states \cite{bose2003quantum,subrahmanyam2004entanglement,shi2005quantum,christandl2005perfect,burgarth2005conclusive,lyakhov2005quantum,bose2007quantum}. Although the maximum distance of the
transfer is limited by the length of the array, it is in principle
possible to realize deterministic state transfer.
\textcolor{black}{Such an approach would be useful to realize scalable
quantum computation in a distributed architecture where quantum computers
with a small size are connected by the qubit array, and there are many
researches along this direction \cite{yao2012scalable,yao2011robust,ping2013practicality}.}
In this paper, we focus on such a quantum state transfer by using the
qubit array.

In the standard schemes, unitary evolution or measurement \textcolor{black}{feed-forwards}
are adopted to perform quantum state transfer in a qubit array.
A sequential implementation of SWAP gates between nearest neighbor
qubits can transfer the quantum state from a node to any other nodes \cite{schuch2003natural,strauch2003quantum,yang2003possible}. Or
combinations of the control-phase gates and measurement \textcolor{black}{feed-forwards} can
teleport the quantum state from one site to another\cite{Raussendorf:2001p368,RBB01a}. However,
these schemes \textcolor{black}{require} accurate operations and/or precise timing for the
implementation of the state transfer. The operational inaccuracy and the
time jittering might accumulates as an error, which could make it
difficult to achieve threshold of the
    topological quantum error correction \cite{raussendorf2007topological,raussendorf2007fault,stephens2013high}.

It is known that dephasing can improve an energy transfer in a qubit-array
under
the effect of inhomogeneous broadening of the qubit frequencies
\cite{engel2007evidence,lee2007coherence,plenio2008dephasing,gilmore2008quantum,rebentrost2009environment,fujii2010anti,uchiyama2018environmental}.
When there is an energy
detuning between the qubits, \textcolor{black}{ and in the absence of time-dependent pulsed operations, the flip-flop interaction between the qubits cannot induce an efficient
energy transfer.}
Interestingly, the existence of dephasing on the qubits
effectively \textcolor{black}{compensates} the energy detuning, and the energy transport can be
enhanced.
However, \textcolor{black}{ when this approach is applied to a one-dimensional system, the energy transfer is bidirectional and the excitations
can be transferred to the right or left of the qubit array}
in a stochastic way.
So the transfer is not as efficient as \textcolor{black}{in} the case of a sequential
implementation of the SWAP gates or quantum teleportation. \textcolor{black}{In these schemes,} the
experimentalist can choose the direction of the energy transfer by changing the
timing, phase, and strength of the pulse sequence.

    \begin{figure*}[t!]
\includegraphics[width=18.0cm, clip]{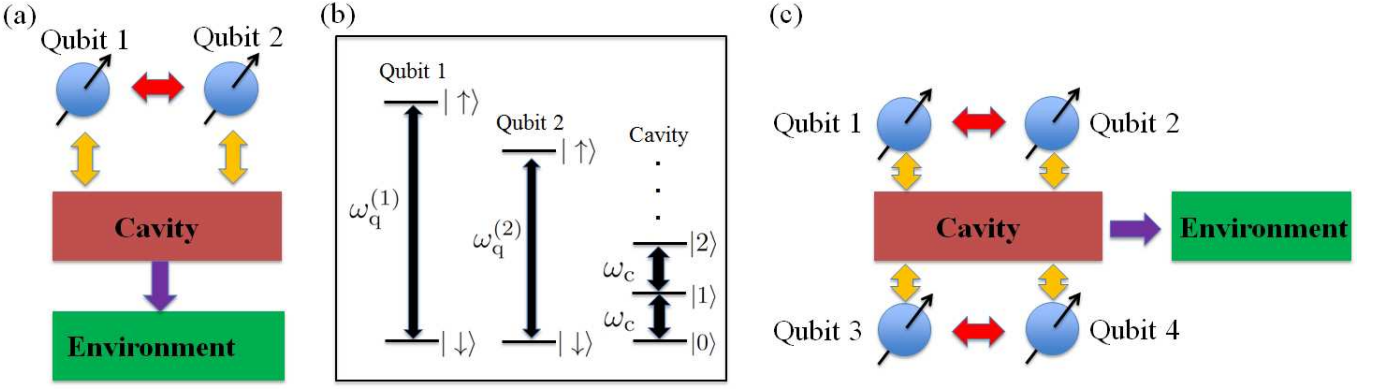}
\caption{
(a) Schematic of our system to implement \textcolor{black}{one-way} energy
     transfer. Two-qubits are collectively coupled with a dissipative
     cavity. An excitation prepared in the qubit 1 can be directionally
     \textcolor{black}{transferred} to the qubit 2.
     (b) An energy diagram of our system when the coupling is absent. The frequency of the qubit 1
     is higher than that of the qubit 2. The cavity
     \textcolor{black}{resonance} frequency is set to
     be around the detuning between the qubits.
     \textcolor{black}{ (c) Schematic of our system to implement a \textcolor{black}{one-way} 
     transfer of quantum states. Four-qubits are collectively coupled with a dissipative
     cavity where the qubit 1 (2) has the same frequency as that of the
     qubit 3 (4). The qubit 1 and qubit 3 are used to prepare an initial
      state described as $\alpha
 |{\uparrow \downarrow} \rangle _{13}+\beta |{\downarrow 
    \uparrow} \rangle _{13}$ where $\alpha $ and $\beta $ denotes
      arbitrary coefficients. In our scheme, an interaction with the dissipative cavity
      can implement a coherence transfer of the prepared state into the qubit 2 and qubit 4.}
     }
\label{dfsexchange}
\end{figure*}

\textcolor{black}{Here, by exploiting decoherence induced by an external environment, we propose a one-way transfer of quantum states, where no active control is needed}
The distinct feature of our scheme is that, after we prepare an initial
state, the transfer from the initial site to the target site
can be implemented automatically by the environment.
 Moreover, due to the non-unitary properties of the decoherence, our
 transfer is \textcolor{black}{one way} so that the quantum state can be \textcolor{black}{transferred along} 
 one direction, and the state remains in the
 target site after the state transfer.

 It is worth mentioning that, by using
 a unitary evolution for the transfer, the quantum state typically
oscillates between different sites, and time-dependent control such as
 turn on/off the interaction is required to make the state stay in the
 target site after the transfer. For example, since a flip-flop
 interaction between two qubits makes an oscillation of the quantum
 state between the two sites, we need to
turn off the interaction to implement a SWAP gate between them where
 time-dependent control is required.
 For practical purposes, automatic
 \textcolor{black}{one-way} transfer might be suitable for quantum information processing
 because \textcolor{black}{it} can avoid potential timing errors that
 typical unitary approaches would suffer from.

 Also, the \textcolor{black}{one-way
 transfer of a quantum state} has a fundamental interest.
 \textcolor{black}{Significant} effort has been made to realize high-performance circulators
and isolators for superconducting quantum circuits \cite{auld1959synthesis,pozar2009microwave,kamal2011noiseless,macklin2015near,barzanjeh2017mechanical}. The non-reciprocal
 properties of these devices
 are useful for protecting quantum states from 
 noise and detecting signals from qubits.
Understanding the mechanism of the non-reciprocal properties of quantum
 systems is important to develop future applications for such 
 directional devices
 \cite{petersen2014chiral,metelmann2015nonreciprocal,sliwa2015reconfigurable,lodahl2017chiral,westig2018josephson,hamann2018nonreciprocity},
 and our results could contributes such a field.

\textcolor{black}{Our} paper is organized as follows. In \textcolor{black}{Sec.}~II, we
consider a two-qubit system collectively coupled \textcolor{black}{to} a dissipative cavity
 to implement the \textcolor{black}{one-way} transfer of an energy excitation from a qubit
 to the other qubit. In this case, the coherence of the initial state is
 not preserved. In \textcolor{black}{Sec.}~III, we describe a four-qubit system
 collectively coupled \textcolor{black}{to} a dissipative cavity. Here, we use \textcolor{black}{two qubits} for
 the preparation of the initial state, and consider the other \textcolor{black}{two qubits} as
 the target sites for the transfer. We show that, by using a
 \textcolor{black}{decoherence-free} subspace \cite{palma1996quantum,beige2000quantum,wu2002creating,altepeter2004experimental}, any single qubit information encoded in a logical
 qubit (composed of two qubits) can be directionally \textcolor{black}{transferred} to the
 target, \textcolor{black}{while the coherence is preserved}. In Sec.~IV, we conclude our
 discussion.

 \section{One-way energy transfer}\label{II}
   \begin{figure*}[htb!]
\includegraphics[width=18.0cm, clip]{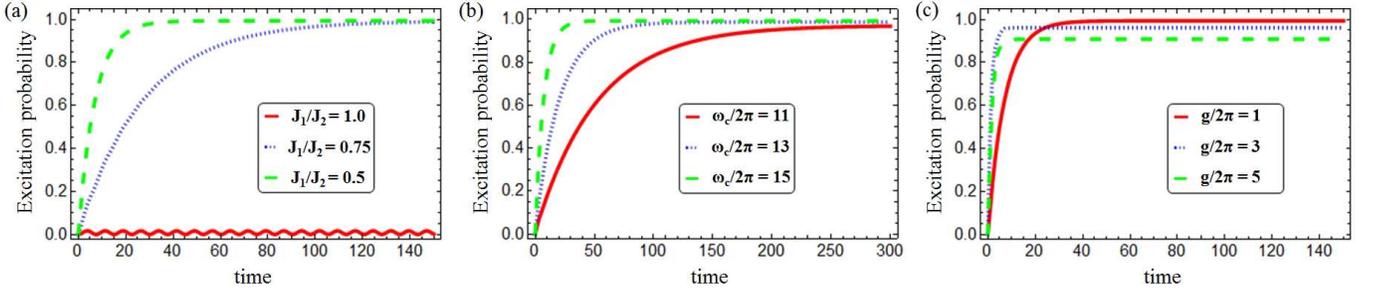}
\caption{
 \textcolor{black}{One-way} energy transfer of our protocol by using two qubits and a cavity. We plot the excitation
  probability $P^{(2)}_{\rm{e}}={\rm{Tr}}[\rho (t) |{\uparrow} \rangle
  _2\langle {\uparrow} |]$ against the time $t$.
    The initial state is $|{\uparrow \downarrow} \rangle _{12}$,
  and so the excitation of the first qubit is directionally \textcolor{black}{transferred}
  to the second qubit.   We set the parameters
  as $\delta \omega _{\rm{q}}^{(1)}/2\pi = 15$, $\delta \omega
  _{\rm{q}}^{(2)}/2\pi = 0$,
    $J_1/2\pi = 2$,
    and $\kappa /2\pi =3$.
(a)Dependency of the cavity-qubit coupling strength
    \textcolor{black}{where we set $g/2\pi = 1$  and $\omega
    _{\rm{c}}/2\pi = 15$.}
    It is worth mentioning that, for $J_1=J_2$, the
    qubit Hamiltonian $H_{\rm{qubit}}$ commutes with the qubit-cavity interaction
    Hamiltonian $H_{\rm{I}}$,
    and so the dissipative cavity
  does not affect the dynamics of the qubit.
    \textcolor{black}{
    (b) Dependency of the cavity frequency where we set
  $J_2/J_1 =0.5$ and $g/2\pi = 1$.  As the cavity frequency becomes closer to the qubit
   detuning $(\delta \omega _{\rm{q}}^{(1)}-\delta \omega _{\rm{q}}^{(2)})$, the transfer speed increases.    
    (c) Dependency of the qubit-qubit coupling strength on
 the excitation
  probability $P^{(2)}_{\rm{e}}={\rm{Tr}}[\rho (t) |{\uparrow} \rangle
  _2\langle {\uparrow} |]$ where we set
  $J_2/J_1 =0.5$ and $\omega _{\rm{c}}/2\pi =15$. As we increase the coupling $g$,  the excitation probability at a steady state
  becomes smaller. 
    }
 }
\label{three}
\end{figure*}
 We describe the model of our system to implement the  \textcolor{black}{one-way}
 energy transfer where the quantum coherence is not preserved. Here, \textcolor{black}{we consider two qubits that interact via flip-flop interaction.}
 There is also a cavity that is
 collectively coupled \textcolor{black}{to} the qubits. The Hamiltonian
 is given
 as follows:\textcolor{black}{
 \begin{eqnarray}
  \hat{H}&=&\hat{H}_{\rm{qubit}}+\hat{H}_{\rm{cavity}}+\hat{H}_{\rm{I}}\nonumber \\
  \hat{H}_{\rm{qubit}}&=&
    \sum_{j=1}^{2}\frac{\hbar \omega ^{(j)}_{\rm{q}} }{2}\hat{\sigma
   }_z^{(j)}
   +\hbar g(\hat{\sigma }_+^{(1)}\hat{\sigma }_-^{(2)}+\hat{\sigma }_-^{(1)}\hat{\sigma }_+^{(2)})
   \nonumber \\
 \hat{H}_{\rm{cavity}}&=&\hbar \omega _{\rm{c}}\hat{a}^{\dagger }\hat{a}
   \nonumber \\
  \hat{H}_{\rm{I}}&=&\hbar (\hat{a}+\hat{a}^{\dagger})\left (\sum_{j=1}^{2}
                                           J_j\hat{\sigma }_z^{(j)}
  \right ) \ , \\
 \end{eqnarray}
 }
 where \textcolor{black}{$\omega ^{(j)}_{\rm{q}}$
 denotes the $j$-th qubit frequencies.}
\textcolor{black}{Also,} \textcolor{black}{ $\omega _{\rm{c}}$  denotes the frequency
 of the cavity,  $g$ is the coupling}
 between the qubits, and $J_j$ denotes the coupling between the qubits and
 cavity. \textcolor{black}{Here, $\hat{\sigma }_+^{(j)}=|{\uparrow}
 \rangle _j \langle {\downarrow} |_j$  and $\hat{\sigma }_-^{(j)}=|{\downarrow} \rangle _j \langle {\uparrow} |_j$ $(j=1,2)$ denote
 the ladder operators}. \textcolor{black}{We set the condition of $\hbar \omega ^{(1)}_{\rm{q}}>\hbar \omega
 ^{(2)}_{\rm{q}}> \omega _{\rm{c}}\gg k_BT$
 throughout of this paper where $k_BT$ denotes the thermal energy
 of the environment. Also, to rescale the qubit frequency, we move into a rotating frame defined by \textcolor{black}{$U=e^{i\frac{ \omega  t}{2}\sum_{j=1}^{2}\hat{\sigma
   }_z^{(j)}}$}, and we obtain the same form of the Hamiltonian except
 that the qubit bare frequency $\omega ^{(j)}_{\rm{q}}$ is replaced by a
 detuning $\delta \omega ^{(j)}_{\rm{q}}=\omega ^{(j)}_{\rm{q}} -\omega $.}
   It is worth mentioning that, our Hamiltonian is different from the
 standard Jaynes Cummings model where the interaction between the qubits and
 cavity is described by
 $(\hat{a}+\hat{a}^{\dagger})(\sum_{j=1}^{2}J_j\hat{\sigma }_x^{(j)})$
 before the rotating wave approximation \cite{JC01a}.
 In our case, since the form of
 the qubit-cavity interaction is
 $(\hat{a}+\hat{a}^{\dagger})(\sum_{j=1}^{2}J_j\hat{\sigma }_z^{(j)})$, our
 Hamiltonian preserves the total number of the excitation of the qubit
 such as $[\textcolor{black}{\hat{H}}, \sum_{j=1}^{2}\hat{\sigma }_z^{(j)}]=0$. This is a
 crucial property to implement our scheme, because otherwise the
 excitation of the qubits could be \textcolor{black}{transferred} into the cavity
  and could be lost from the qubits.
 Moreover, in our scheme, the cavity is supposed to be coupled with an environment that
 induces a photon loss.  We assume that the thermal energy of the environment is
 much smaller than the frequency of the cavity. This means that the cavity emits
 the photon to the environment but the cavity does not absorb the photon
 from the environment \cite{gardiner1991quantum}.
 In order to include such a photon loss, we 
 adopt the following Lindblad master equation in the rotating frame:
 \textcolor{black}{
 \begin{eqnarray}
  \frac{d \hat{\rho} (t)}{dt}=-\frac{i}{\hbar }[\hat{H},\hat{\rho}(t) ]-\frac{\kappa }{2}(\hat{a}^{\dagger
   }\hat{a}\hat{\rho}(t) +\hat{\rho}(t) \hat{a}^{\dagger }\hat{a}-2\hat{a}\hat{\rho}(t)
   \hat{a}^{\dagger })\label{lind} \ ,\ \ 
 \end{eqnarray}
 }
 where $\kappa$ denotes the photon loss rate.

 We solve the Lindblad master equation \textcolor{black}{given by}
 Eq. (\ref{lind}), and
 \textcolor{black}{illustrate its energy transfer behavior.}
 \textcolor{black}{From now on we are going to use the simplified
 notation for the states $|\sigma\rangle _{i}\otimes |\tilde{\sigma}
 \rangle _{j}=|\sigma \tilde{\sigma} \rangle _{i j}$, where
 $\sigma, \tilde{\sigma}=\uparrow,\downarrow$.}
 We set the initial state \textcolor{black}{$\hat{\rho}(0)=|\Psi(0) \rangle
\langle \Psi(0)|$, where 
 $|\Psi(0)\rangle=|{\uparrow \downarrow} \rangle _{12}$. Now we proceed to
 investigate the evolution of the initial state. With this aim,
 we plot the excitation probability} of the second qubit \textcolor{black}{$P^{(2)}_{\rm{e}}={\rm{Tr}}[\rho (t) |{\uparrow }\rangle
  _2\langle{{\uparrow}} |]$} in the Fig. \ref{three}.
  As long as $J_1 > J_2$, the excitation
 probability almost monotonically increases, and converges to a steady state. 
 In this parameter
 regime, the excitation probability for the steady state
 becomes more than $99\%$, and this shows a high
 efficiency of our one-way transfer scheme.
 Also, we investigate how the cavity frequency affects the energy transfer.
 As the cavity frequency becomes closer to the qubit frequency difference
 $(\omega _{\rm{q}}^{(1)}- \omega _{\rm{q}}^{(2)})$, the transfer becomes faster as shown in the
 Fig. \ref{three}.
 On the other hand,
 a larger qubit-qubit coupling
 strength makes the excitation probability at the target qubit smaller
 when the system reaches the steady state as shown in the Fig. \ref{three}.

 We will describe the reasons why our scheme can achieve
 almost the unit efficiency of one-way energy transfer as shown in the Fig. \ref{three}. For
 this purpose, we diagonalize $\hat{H}_{\rm{qubit}}=\sum_{j=1}^{4}E_j |E_j\rangle \langle
   E_j|$. \textcolor{black}{
   The
 Hamiltonian preserves the total number of the excitation of the qubit
 such as $[\hat{H}_{\rm{qubit}}, \sum_{j=1}^{2}\hat{\sigma
 }_z^{(j)}]=0$, and so we can analytically obtain the eigenvectors as follows:}
   \textcolor{black}{
 \begin{eqnarray}
  |E_1\rangle &=&|{\uparrow \uparrow}  \rangle_{12}  \label{e1vec} \\
  |E_2\rangle &=&\cos  \frac{\theta }{2}  |{\uparrow \downarrow} \rangle_{12}  +
   \sin \frac{\theta }{2}|{\downarrow \uparrow} \rangle_{12}   \label{e2vec} \\
    |E_3\rangle &=&\sin \frac{\theta }{2} |{\uparrow \downarrow} \rangle_{12}  -
   \cos \frac{\theta }{2} |{\downarrow \uparrow} \rangle_{12}   \label{e3vec} \\
  |E_4\rangle &=&|{\downarrow \downarrow}  \rangle_{12}   \label{e4vec}  \ , 
 \end{eqnarray}
 }
 where we have
 \begin{eqnarray}
 && \cos \theta =\frac{\delta \omega _{\rm{q}}^{(1)}-\delta \omega
 _{\rm{q}}^{(2)}}{\sqrt{(\delta \omega _{\rm{q}}^{(1)}-\delta \omega
 _{\rm{q}}^{(2)})^2+4g^2}}\nonumber \\
  &&\sin \theta =\frac{2g}{\sqrt{(\delta
 \omega _{\rm{q}}^{(1)}-\delta \omega
 _{\rm{q}}^{(2)})^2+4g^2}}\nonumber \ .
 \end{eqnarray}
 In our paper, we assume $|\delta \omega _{\rm{q}}^{(1)}-\delta \omega _{\rm{q}}^{(2)}|\gg g>0$, and so we have
 $\cos \frac{\theta }{2}\simeq 1$ and 
 $\sin \frac{\theta }{2}\simeq \frac{\theta }{2}\ll 1$.
 
 Also, the eigenvalues are as follows:
 \begin{eqnarray}
    E_1 &=&\frac{\hbar \delta \omega _{\rm{q}}^{(1)}+\hbar \delta \omega _{\rm{q}}^{(2)}}{2}
   \label{e1} \\
  E_2&=&\frac{\hbar \sqrt{(\delta \omega _{\rm{q}}^{(1)}-\delta \omega _{\rm{q}}^{(2)})^2+4g^2}}{2} \label{e2} \\
  E_3&=&-\frac{\hbar \sqrt{(\delta \omega _{\rm{q}}^{(1)}-\delta \omega _{\rm{q}}^{(2)})^2+4g^2}}{2} \label{e3} \\
  E_4&=&-\frac{\hbar \delta \omega _{\rm{q}}^{(1)}+\hbar \delta \omega _{\rm{q}}^{(2)}}{2} \label{e4}
 \ .
 \end{eqnarray}
 It is worth mentioning that, if we have a uniform interaction between
 the qubits and cavity such as $J_1=J_2$,
the interaction Hamiltonian $H_{\rm{I}}$ commutes with $H_{\rm{qubit}}$, and the
 cavity does not induce the energy transfer as shown in the
 Fig. \ref{three}.
 So we consider a case of an inhomogeneous coupling ($J_1\neq J_2$) with the cavity.
 Importantly, due to the small but a finite effect of the coupling
 strength $g$, the interaction from the cavity induces a transition between
 $|E_2\rangle $ and $|E_3\rangle $ where we have
 $\langle
 E_2|H_{\rm{I}}|E_3\rangle \neq 0$.
 This means that,
 when the resonant
 frequency $\omega _{\rm{c}}$ of the cavity \textcolor{black}{is closer to}
 the energy
 difference between $|E_2\rangle $ and $|E_3\rangle $, there is an \textcolor{black}{efficient}
 energy exchange between the cavity and qubits. Actually, this explains
 why the transfer becomes more efficient as the cavity frequency becomes
 closer to the qubit-qubit detuning $(\delta \omega
 _{\rm{q}}^{(1)}-\delta \omega _{\rm{q}}^{(2)})$
 in the Fig. \ref{three}. 
Moreover, due to the strong coupling of the cavity with the
 low-temperature environment, the
 cavity emits a photon immediately after catching the excitation from
 the qubit, and the cavity can approximately stay in a vacuum state.
 Therefore, the initial state
 $|E_2\rangle $ will irreversibly evolve into a steady state $|E_3\rangle $ in our
 system.
 On the other hand, although the energy of
 $|E_4\rangle $ is lower than that of $|E_3\rangle $, a transition from $|E_3\rangle $ to
 $|E_4\rangle $ is prohibited due to a zero transition matrix element of $\langle
 E_3|H_{\rm{I}}|E_4\rangle = 0$. 
 It is worth mentioning that, as the qubit-qubit coupling $g$ becomes
 larger, a deviation of the state $|E_2\rangle $ ($|E_3\rangle $)  from $|{\uparrow
 \downarrow} \rangle _{12}$ ($|{\downarrow \uparrow }\rangle _{12}$) becomes
 larger. With a large $g$, the irreversible transition from $|E_2\rangle $
 to $|E_3\rangle $
 does not correspond to our desired transition (from $|{\uparrow \downarrow}
 \rangle _{12}$ to $|{\downarrow \uparrow} \rangle _{12}$), which explains the
 reason why the excitation of the target qubit in the steady state
 becomes smaller
 with a larger coupling $g$ in the Fig. \ref{three}.

 \section{One-way quantum state transfer}
\begin{figure}[h!]
\includegraphics[width=8.0cm, clip]{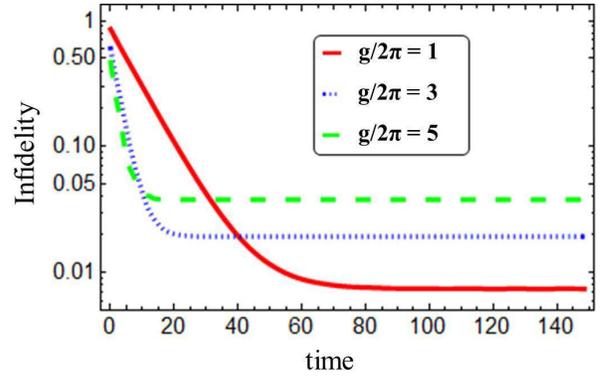}
\caption{
\textcolor{black}{One-way} quantum state transfer of our protocol by using four qubits and a
    cavity. The initial state is $\frac{1}{\sqrt{2}}(|{\uparrow \downarrow} \rangle _{13}+|{\downarrow
    \uparrow} \rangle _{13}) |{\downarrow \downarrow} \rangle _{24}$. 
    We define a fidelity with the target state $F={\rm{Tr}}[\rho
    (t) |\psi _{\rm{Bell}}\rangle \langle \psi _{\rm{Bell}}|]$ where
    $|\psi _{\rm{Bell}}\rangle =|{\downarrow \downarrow}
    \rangle _{13}\frac{1}{\sqrt{2}}(|{\uparrow \downarrow} \rangle _{24}+|{\downarrow
    \uparrow} \rangle _{24})$. We plot the infidelity $1-F$ in a
    log scale against a time $t$. As we decreases $g$, the infidelity
    becomes smaller at the steady state, which is consistent with the
    results in the Fig. \ref{three}. We set
  $J_2/J_1 =0.5$, $J_1=J_3$, and $J_2=J_4$.
    We use the same
  parameters as that used in the Fig. \ref{three} (c)
    except the
  qubit-qubit coupling.
 }
\label{dfsbell}
\end{figure}

 \textcolor{black}{Let us now} describe the model of our system to implement a \textcolor{black}{one-way} coherent quantum
 state transfer. In the scheme explained in the Sec. \ref{II},
 the decoherence from the environment will destroy the quantum
 coherence, and so only the energy of the initial state can be
 \textcolor{black}{transferred} to the target qubit. On the other hand, in this section, we will show that
a quantum state can be directionally \textcolor{black}{transferred} in a qubit array.
 Here, we consider four qubits collectively coupled with a dissipative cavity, as shown in the Fig. \ref{dfsexchange}.
 The Hamiltonian in the rotating frame is given as follows.
 \textcolor{black}{
 \begin{eqnarray}
  \hat{H}_{\rm{QT}}&=& \hat{H}_{\rm{qubit}}+ \hat{H}_{\rm{cavity}}+ \hat{H}_{\rm{I}}\
   \nonumber \\
   \hat{H}_{\rm{qubit}}&=&\sum_{j=1}^{4}\frac{\hbar \delta \omega ^{(j)}_{\rm{q}} }{2}\hat{\sigma
   }_z^{(j)}  + \hbar g\sum_{j=1,3}(\hat{\sigma }_+^{(j)}\hat{\sigma
   }_-^{(j+1)}+\hat{\sigma }_-^{(j)}
   \hat{\sigma }_+^{(j+1)}) 
   \nonumber \\
   \hat{H}_{\rm{cavity}}&=&\hbar \omega _{\rm{c}}\hat{a}^{\dagger }\hat{a} \ \ \
  \nonumber \\
   \hat{H}_{\rm{I}}&=&\hbar(\hat{a}+\hat{a}^{\dagger})\left ( \sum_{j=1}^{4}
   J_j\hat{\sigma }_z^{(j)} \right ) \ ,
 \end{eqnarray}
 }
where the first (third) qubit has an energy-exchange interaction with the second
(fourth) qubit. We assume $\delta \omega _{\rm{q}}^{(1)} > \delta \omega
_{\rm{q}}^{(2)}$ ($\delta \omega _{\rm{q}}^{(3)}> \delta \omega _{\rm{q}}^{(4)}$).
Also, we assume that the qubit 1 (2) and qubit 3 (4) has the
same parameter such as $\delta \omega _{\rm{q}}^{(1)}=\delta \omega _{\rm{q}}^{(3)}$,
$\delta \omega _{\rm{q}}^{(2)}=\delta \omega _{\rm{q}}^{(4)}$, $J_1=J_3$, and
$J_2=J_4$.
The key idea is to use a decoherence free subspace \cite{palma1996quantum,beige2000quantum,wu2002creating,altepeter2004experimental}.
By using the qubits 1 and 3, we can define a logical qubit
$|0_{\rm{L}}\rangle _{13} =|{\uparrow
\downarrow} \rangle _{13}$ and $|1_{\rm{L}}\rangle _{13}=|{\downarrow
\uparrow} \rangle _{13}$. We can also define such a logical qubit by using the
qubit 2 and 4. Since the qubit $1$ ($2$) and $3$ ($4$) are identical for
the dissipative cavity, the environment cannot distinguish them, which
makes it possible to implement a \textcolor{black}{one-way} transfer of quantum states as
we will describe later.
Similar to the case of the energy transfer in the Sec.
\ref{II}, we consider a photon loss of the cavity, and so
we will
 solve the following Lindblad master equation:
 \textcolor{black}{
 \begin{eqnarray}
  \frac{d\hat{\rho} (t)}{dt}=-\frac{i}{\hbar}[\hat{H}_{\rm{QT}},\hat{\rho}(t) ]-\frac{\kappa }{2}(\hat{a}^{\dagger
   }\hat{a}\hat{\rho}(t) +\hat{\rho}(t) \hat{a}^{\dagger }\hat{a}-2\hat{a}\hat{\rho}(t)
   \hat{a}^{\dagger }).
   \nonumber
   \label{lindqt}
 \end{eqnarray}
 }
We show the efficiency of our scheme to transfer the quantum state  by
solving the Lindblad master equation.
\textcolor{black}{Firstly, we set} the initial state as
$\frac{1}{\sqrt{2}}(|{\uparrow \downarrow} \rangle _{13}+|{\downarrow 
    \uparrow} \rangle _{13}) |{\downarrow \downarrow} \rangle _{24}$. 
    Also, we define a fidelity with the target state \textcolor{black}{$F={\rm{Tr}}[\hat{\rho}
    (t) |\psi _{\rm{Bell}}\rangle \langle \psi _{\rm{Bell}}|]$} where
    $|\psi _{\rm{Bell}}\rangle =|{\downarrow \downarrow}
    \rangle _{13}\frac{1}{\sqrt{2}}(|{\uparrow \downarrow} \rangle _{24}+|{\downarrow
    \uparrow} \rangle _{24})$. In the Fig. \ref{dfsbell}, we plot the infidelity $1-F$ 
    against a time $t$.
    By choosing the parameters in the Fig. \ref{dfsbell}, the fidelity
    can be more than $99\%$, which is more than the threshold of the
    topological quantum error correction \cite{raussendorf2007topological,raussendorf2007fault,stephens2013high}. 
 \textcolor{black}{Secondly}, we numerically confirm that, for \textcolor{black}{several} other initial states
 described as $(\alpha
 |{\uparrow \downarrow} \rangle _{13}+\beta |{\downarrow 
    \uparrow} \rangle _{13}) |{\downarrow \downarrow} \rangle _{24}$ where
    $\alpha $ and $\beta $ are coefficients, we
    can obtain a similar fidelity by using the same parameters.
    So these show that we can implement a directional transfer of 1
    qubit information
    by using four qubits and a dissipative cavity.

We explain the reasons why we can send the quantum state in a
directional way. 
Since the qubit 1 (2) and 3(4) are identical,
the dissipative cavity cannot
distinguish whether the excitation is located at the qubit 1 (2) or at
the qubit 3 (4). This means that the coherence between
$|0_{\rm{L}}\rangle _{13}$ ($|0_{\rm{L}}\rangle _{24}$)
 and $|1_{\rm{L}}\rangle _{13}$ ($|1_{\rm{L}}\rangle _{24}$) is
 preserved in our scheme.
 On the other hand, as long as we set $J_1 > J_2$, the dissipative
 cavity still can distinguish whether the excitation is on the qubit 1
 or on the qubit 2, which means that the directional excitation transfer
 can be performed similar to the mechanism described in the Sec.
 \ref{II}.
 Similar to the Eqs. (\ref{e1vec})-(\ref{e4}), we can diagonalize
 $H_{\rm{qubit}}$. The second excited state 
 and the third excited state are approximated as
 \textcolor{black}{$|{\downarrow
\uparrow} \rangle _{12}$ ($|{\downarrow
\uparrow} \rangle _{34}$) and $|{\uparrow
\downarrow} \rangle _{12}$ ($|{\uparrow
\downarrow} \rangle _{34}$)} for  $H_{\rm{qubit}}$
 where the dissipative cavity can induce an
irreversible transition between them as described in the Sec \ref{II}, and this approximation becomes better for
$(\delta \omega _{\rm{q}}^{(1)}-\delta \omega _{\rm{q}}^{(2)})\gg g$
($(\delta \omega _{\rm{q}}^{(3)}-\delta \omega _{\rm{q}}^{(4)})\gg g$).
 This explains why the infidelity can be smaller for a smaller $g$ in the
 Fig. \ref{dfsbell}.

  \section{Conclusion}\label{IV}
  In conclusion, we \textcolor{black}{have shown} a scheme to implement a directional transfer of
  a quantum state by using a tailored environment. Since we use
  decoherence for the transfer, this is an irreversible
  process. Moreover, once we
  prepare an initial state, the transfer can be automatically
  implemented without any time dependent operations. 
  To implement a one-way energy transfer, two qubits and a dissipative
  cavity are required, while we need four qubits and a dissipative cavity
  for a quantum state transfer with preserved coherence.
  Our scheme provides not only an alternative way to implement a quantum
  state transfer for quantum information
  processing but also a deep understanding of a non-reciprocal device
  for future superconducting circuit applications. 

  \section*{ACKNOWLEDGMENTS}
  We thank Keisuke Fujii, Emi Yukawa, and Ivan Iakoupov for helpful discussions.
This work was supported
by CREST (JPMJCR1774), JST,
and in part by
MEXT Grants-in-Aid for Scientific Research on Innovative Areas ``Science of
hybrid quantum systems'' (Grant No. 15H05870).

Note added - While we are preparing our manuscript,
a related paper appeared on arXiv, which shows another way to perform
the directional quantum state transfer by dissipation \cite{wang2018directional}. 
  

\end{document}